\documentclass[preprint,12pt]{elsarticle}

\usepackage{amssymb} 
\usepackage{amsmath} 
\usepackage{hyperref}

\newcommand{\bra}[1]{\langle #1 \vert}
\newcommand{\ket}[1]{\vert #1 \rangle}

\journal{}

\begin{document}

\begin{frontmatter}

\title{Localization properties and high-fidelity state transfer in electronic hopping models with correlated disorder}

\author{G. M. A. Almeida, C.V.C. Mendes, M. L. Lyra, and  F. A. B. F. de Moura}

\address{Instituto de  F\'{\i}sica, Universidade Federal de
Alagoas, Macei\'{o} AL 57072-970, Brazil}

\begin{abstract}
We investigate a tight-binding electronic chain
featuring diagonal and off-diagonal disorder, these being modelled through the long-range-correlated fractional Brownian motion.
Particularly, by employing exact diagonalization methods, we evaluate how the eigenstate spectrum of the system
and its related single-particle dynamics respond to 
both competing sources of disorder.
Moreover, we report the possibility of carrying out efficient end-to-end quantum-state transfer protocols 
even in the presence of such generalized disorder due to the appearance of extended states around the middle of the band 
in the limit of strong correlations.  
\end{abstract}  
\begin{keyword} 
diffusive spreading, correlated disorder, localization, quantum-state transfer
\end{keyword}

\end{frontmatter}

\section{Introduction}

In the past few decades, there has been a growing interest in investigating quantum transport properties 
of low dimensional disordered lattices~\cite{abrahams,kramer,epl2016,aop2016,prl2008,prb2008,njp1,jpcm2009,prl2009,prl1,
garreau2011,prb2011,jacob2011,jpcm1,jpcm2,jpcm3},
most of them based on Anderson scaling theory.  
In general lines, it is well established that there are no extended eigenstates in low-dimensional systems for any amount of uncorrelated disorder. 
The breakdown of standard Anderson localization theory 
was put forward
about thirty years ago by Flores and Dunpap \cite{flores,dunlap}.  
They pointed out that the presence of short-range correlations in the disorder distribution yielded 
the appearance of extended states in the spectrum of disordered chains. 
That could explain to a great extent some unusual transport properties of several types of polymers \cite{flores,dunlap}.
%
Right after this discovery, a handful of works came along to investigate the role of disorder correlations, either short- or long-ranged, in 
wide variety of physical systems~\cite{chico,chico2,izrailev,mauricio2d,prb2006,chico2d2007,prl2003,chico2010,liu,xiong1,
Bellani99,Bellani,shima,thaila2011,hop,fehske,preport1,prbcorr1,prbcorr2,greg,prb2017, almeida17-1}. 
%
Particularly, it was shown in Refs. \cite{chico,izrailev} that 
long-range correlated random potentials can actually allow for mobility edges in 1D disordered models. 
%
In Ref.~\cite{chico}, 
that specific kind of fluctuations was generated using
the trace of a fractional Brownian motion whose intrinsic correlations decay following a power law. 
Through numerical renormalization methods, 
it was show that this model exhibits a phase of extended states around the center of the band~\cite{chico}. 
Tackling the same problem, the authors in ~\cite{izrailev} 
applied a analytical perturbation technique and came up with a direct relationship 
between the localization length and the characteristics of the intrinsic correlations in the disorder distribution. 
A few years later, the above results were validated through experiments carried out 
in microwave  guides featuring correlated scatters~\cite{apl2}. 
The authors demonstrated that intrinsic long-range correlations within the scatters distribution ultimately 
improve the wave transmission. On the theoretical side, the Anderson model with
long-range correlated hopping fluctuations (off-diagonal disorder)  was studied in Refs.~\cite{chico2,thaila2011}. Likewise, it was found that 
strong correlations promote the appearance of a phase of extended states close to the center of the band. 

In this work we provide further progress along those lines. In particular, we consider two sources of disorder acting simultaneously
on the potentials as well as on the hopping strengths of the chain, both exhibiting long-range correlated fluctuations
generated by the fractional Brownian motion. This model embodies a generalized disordered scenario
which we aim to push on its capability of supporting extended states in the middle of the band 
thereby weakening Anderson localization. By looking at the participation ratio of eigenstates and also
at the dynamics of the system through its mean square displacement for an delta-like initial state
we find out the chain allows for propagating modes if substantial long-range correlations
are taking place in both sources of disorder.
Looking forward possible applications in the field of quantum-information processing, we also 
investigate whether such a model of generalized disorder would allow for 
realizing weak-coupling quantum-state transfer protocols \cite{bose03, wojcik05, wojcik07, almeida16}.
%
The point is that when designing chains for transmitting quantum states from one point to another -- which is a crucial requirement in quantum networks \cite{cirac97} --
one should take into account the possibility of 
undesired fluctuations taking place due to experimental errors \cite{almeida17-1,dechiara05, burgarth05, tsomokos07, yao11, zwick11, ashhab15, estarellas17}, 
that including correlated noise \cite{almeida17-1,dechiara05, burgarth05}. Our calculations reveal that an electron (or a properly encoded qubit) 
can be almost fully transferred through the noisy bulk of the chain depending upon specific sets of parameters.

\section{Model and Formalism}

We consider a $N$-site linear chain described by the electronic tight-binding Hamiltonian ($\hbar = 1$)
\begin{eqnarray}
\label{hamilcompleto}
H & = & \sum_{n=1}^{N}\epsilon_n|n\rangle\langle n| + \sum_{n=1}^{N-1}J_{n}(|n\rangle\langle n + 1|+\mathrm{h.c.}),
\end{eqnarray}
written in the Wannier basis set $\lbrace \ket{n} \rbrace$ accounting for the electron position, where
$\epsilon_n$ is the on-site potential and $J_n$ is the hopping strength, those being the source of static disorder.
Those parameters are here expressed in terms of energy unit $J\equiv 1$.
Specifically, we assume that both quantities fluctuate such that their corresponding disorder distributions
come with intrinsic long-range correlations modelled via the 
fractional Brownian motion~\cite{chico,mauricio2d,chico2d2007,prl2003}   
\begin{equation}
\epsilon_{n}, J_{n} = \sum_{k=1}^{N/2}\frac{1}{k^{\gamma/2}}\cos\Bigg(\frac{2\pi nk}{N}+\phi_k\Bigg).\ 
\label{mbf}
\end{equation}
We emphasize that the sequence generated by the equation above exhibits a power-law spectrum 
$~1/k^{\gamma}$ and $\phi_{k}$ represents a random phase uniformly distributed within the range $[0,2\pi]$. 
For $\gamma = 0$, the sequence is fairly uncorrelated. On the other hand,  $\gamma > 0$ brings about long-range
correlations in the disorder sequence. Therefore, exponent $\gamma$ 
stand out as very important parameter in our work since it controls the  \textit{degree} of correlations 
within the disordered sequence. 
Hereafter, Eq. (\ref{mbf}) will be used for generating disorder distributions for both $\epsilon_n$ and $J_n$ but with a few remarks:
(i) for $\epsilon_n$  we attribute  $\gamma \rightarrow \alpha$ and
normalize the entire sequence so that $\langle\epsilon_n \rangle=0$ and $\langle \epsilon_n^2\rangle=1$; (ii) for $J_n$ 
we set $\gamma \rightarrow \beta$ and redefine $J_n \rightarrow \tanh{(J_n)}+2$ after normalization in order to rule out
possible null hopping strengths.
It is also important to note that each sequence for $\epsilon_n$ and $J_n$ is generated using distinct sets of phases, $\lbrace\phi_k \rbrace$. 
In summary, our model contains two \textit{independent} parameters  $\alpha$ and $\beta$
that account for the degree of correlations for both diagonal and off-diagonal sources of disorder.
 
Our quantities of interest are all obtained through exact diagonalization of Hamiltonian (\ref{hamilcompleto}) which gives
us the eigenvalues $\{E_j\}$ and its corresponding eigenvectors $\ket{\psi^j}=f_n^j\ket{n}$. 
Our first task will be evaluating the participation ratio defined as~\cite{chico2d2007}
\begin{eqnarray}
\xi^j =\frac{1}{\sum_n |f_n^j|^4}.
\label{participacao}
\end{eqnarray}
This measure provides an estimate of the number of bare states a given eigenstate is spread on, i.e., it quantifies
the degree of localization.
In particular, the 
participation number becomes size-independent for localized wave-packets and diverges with $N$ for extended ones.
In addition, we investigate the electronic time evolution through the chain. We initialize the initial wave-packet 
in $\ket{\psi(0)}=\sum_n c_n(0)\ket{n}$ where $c_n(0)=\delta_{n,n_0}$. The electronic state at time $t$ can thus be obtained 
from 
$\ket{\psi(t)}=\sum_n c_n(t)\ket{n}=e^{-iHt}\ket{\psi(0)}$,
where 
\begin{equation} \label{fnt}
c_n(t)=\sum_j f_{n_0}^j f_n^je^{-iE_jt}.
\end{equation}
By using the relations above we can compute the width $\sigma$ of the electronic wave-packet through~\cite{moura_phye_2012}
\begin{eqnarray}
\sigma(t)=\sqrt{\sum_n (n-\langle n(t)\rangle)^2|c_n(t)|^2},
\label{larguramedia}
\end{eqnarray}
where  $\langle n(t)\rangle=\sum_n n |c_n(t)|^{2}$ is the electronic  average position. 
Note that $\sigma (t)$ goes from $0$, for a wave function confined to
a single site, to $O(N)$ for a wave extended over the
whole system. Note that we can also compute the time-dependent participation number defined as $\xi(t)=1/\sum_n |c_n(t)|^4$. Both quantities are distinct ways  to obtain  an estimate of the size of the  wave-packet  at time $t$~\cite{chico2d2007,moura_phye_2012}.

\section{Results}

After having introduced our main tools in the previous section it is now time to investigate the actual role 
of diagonal and off-diagonal sources of disorder 
acting simultaneously in the chain.

\subsection{Localization properties}

\begin{figure}[t!]
\begin{center}
\includegraphics[width=0.99\textwidth]{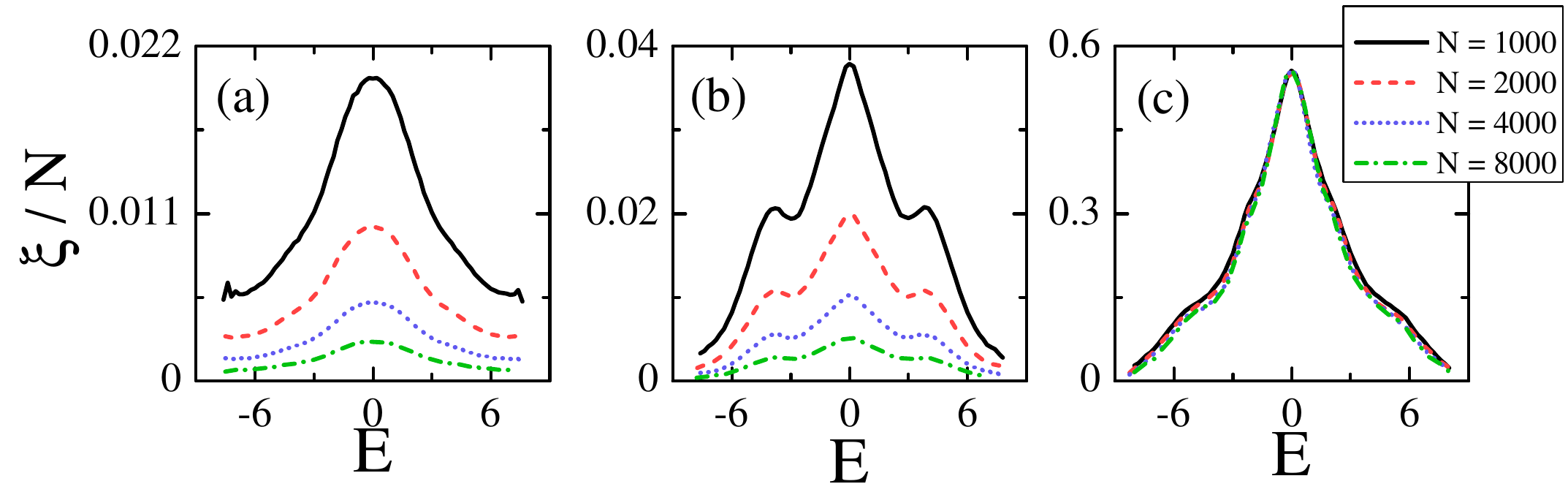}
\caption{Rescaled participation number  $\xi/N$ versus $E$  for (a) $\alpha=3$, $\beta=0$; (b) $\alpha=0$, $\beta=3$; and (c) $\alpha=3$, $\beta=3$ for several system sizes. Observe that the center of the band tends to build up delocalized-like states.}
\label{fig1}
\end{center}
\end{figure}

\begin{figure}[t!]
\begin{center}
\includegraphics[width=0.5\textwidth]{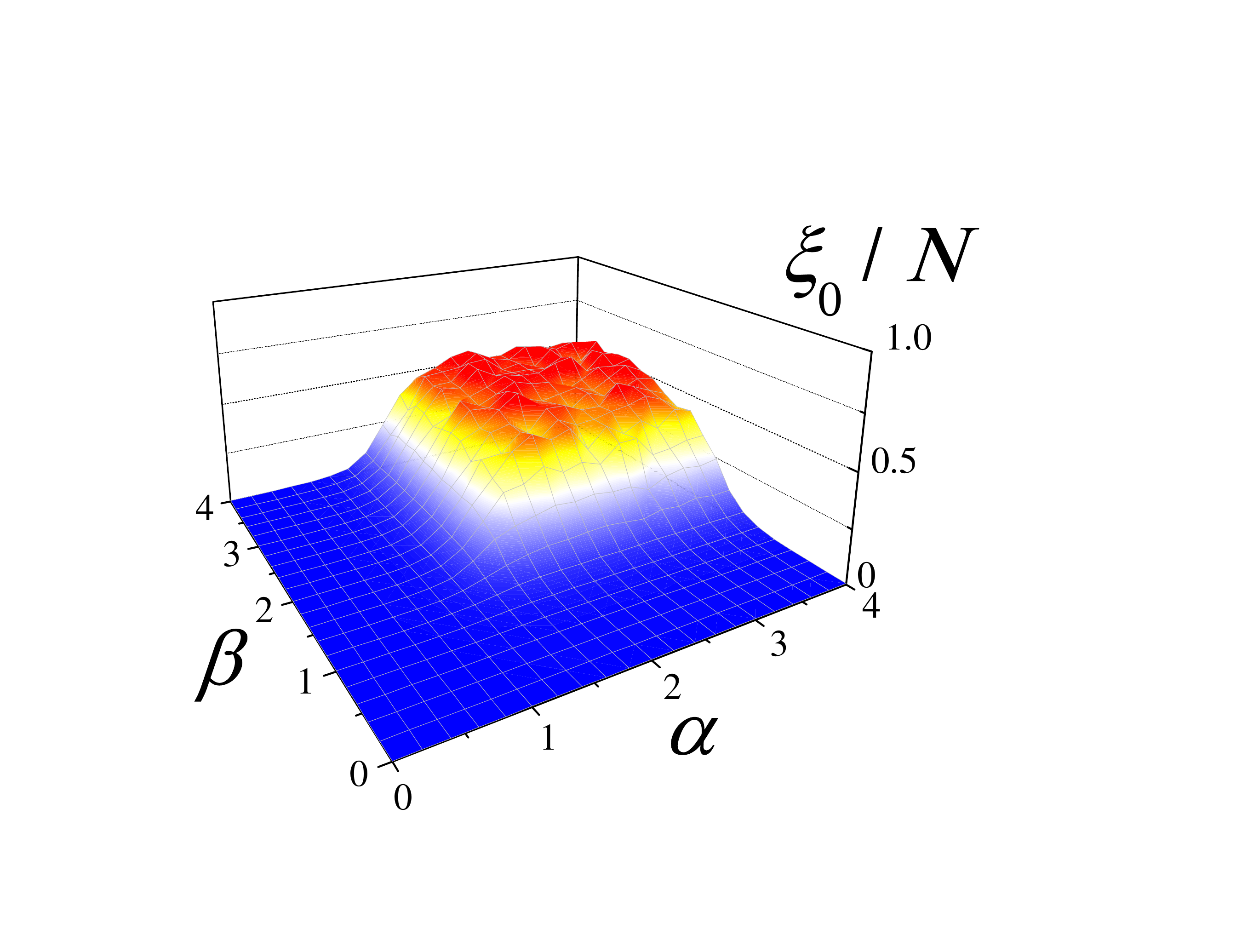}
\caption{Rescaled participation number around the band center, $\xi(E\approx 0)/N=\xi_0/N$, versus $\alpha$ and $\beta$ for $N=8000$ sites. The chain is able to support extended states around the center of the band only when both diagonal and off-diagonal sources of disorder are, each, long-range correlated obeying, roughly, $\alpha, \beta > 2$.}
\label{fig2}
\end{center}
\end{figure}

We start our analysis showing results for the participation ratio of the entire eigenstates set. 
It should be emphasized that every quantity evaluated in this work was properly averaged
over many distinct realizations of disorder. 
%
The total number of eigenstates  $N_E=NM$ was larger than $10^5$ for all calculations, $M$ being the number of samples.
We averaged $\xi^j$ over a small window around energy $E$ and therefore we 
are looking towards the quantity $\xi(E)=(\sum_{E_j>E-\Delta E}^{E_j<E+\Delta E}\xi^j)/n(E)$, where $n(E)$ is the number of eigenvalues $\{E_j\}$ within the interval $[E-\Delta E,E+\Delta E]$. 
Herein we fix $\Delta E=0.2$. 

In Fig.~\ref{fig1} we plot the rescaled mean participation number $\xi/N$ versus energy $E$ for 
many combinations of $\alpha$ and $\beta$.
Calculations were done for $N=1000$ up to $8000$ sites. 
We observe in Figs. \ref{fig1} (a) and \ref{fig1}(b) that 
$\xi/N$ decreases as the system size $N$ is increased regardless of the $E$ value. This is
a clear signature that all eigenstates become localized at the thermodynamic limit. 
On the other hand, Fig. \ref{fig1}(c) reveals a rather interesting behavior.
Close to the band center, the rescaled participation number remains constant thus indicating the appearance of extended states at this region.  For $|E|>>0$ we observe that $\xi/N$ decreases with $N$ 
what indicates the presence of localized states far from the band center. 
Thereby, our calculations show that one-dimensional systems
featuring both diagonal and off-diagonal disorder only display 
extended states whenever both sources of fluctuations are augmented with strong long-range correlations.  
If only either $\alpha$ \textit{or} $\beta$ is greater than zero,
the electron transport can be suppressed by  the presence of uncorrelated randomness in the lattice. 
We can further observe this feature by analyzing  Fig.~\ref{fig2} where we plot the mean participation number around the band center  $\xi_0/N\equiv \xi(E\approx 0)/N$ versus $\alpha$ and $\beta$ for $N=8000$. 
We note that only for $\alpha$ \textit{and} $\beta$ larger than $2$ we 
are to obtain the rescaled participation number $\xi_0/N \approx 0.58(2)$ which is very close to the
corresponding value of extended states in ordered chains with open-boundary conditions, that is $2/3$.
Our outcomes are also in agreement with
the rescaled participation number for extended states in disordered systems found elsewhere~\cite{mauricio2d,chico2d2007, thaila2011}.  

Furthermore, it is relevant to point out that, generally speaking, 
$\gamma$ is 
is related to the so-called Hurst exponent $\mathcal{H}$
through $\mathcal{H} = (\gamma - 1)/2$ which describes
the long-term memory
of a given series.  The set spanned by Eq. (\ref{mbf})  is said to be nonstationary when $\gamma > 1$ 
and persistent (anti-persistent) when $\gamma > 2$ ($\gamma < 2$).  When $\alpha = 2$
the series corresponds exactly to the trace
of the Brownian motion. 
Moreover, as shown in \cite{chico} in the case of on-site disorder only, $\alpha= 2$ 
marks the transition point between 
Anderson-like insulator and metallic phases with sharp mobility edges.

\subsection{Time dynamics and quantum-state transfer}
 
The interplay between localized and delocalized states we have seen in the previous section
allows for a rich variety of dynamical regimes \cite{prl2003}. Our goal now is 
explore how the competition between two independent sources of correlated disorder
reflects upon the spreading profile of the initial state of a single electron. Right after that we will tackle
a very appealing application of such platforms in the context of quantum information processing. 
 
\begin{figure}[t!]
\begin{center}
\includegraphics[width=0.98\textwidth]{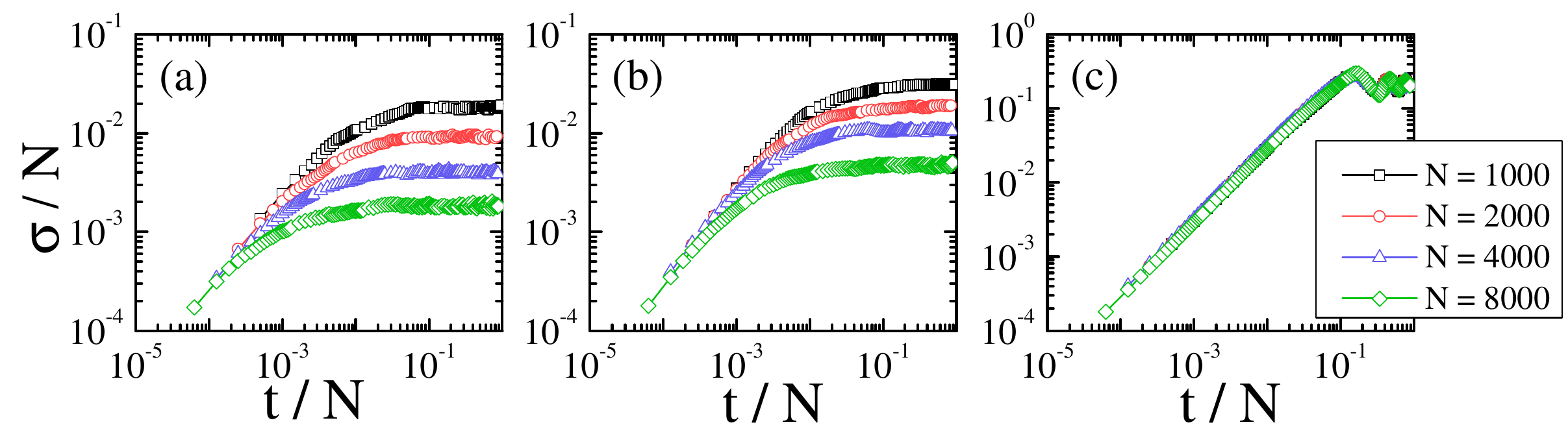}
\caption{Rescaled square root of the mean square displacement ($\sigma/N$) versus rescaled time ($t/N$)  for (a) $\alpha=3$, $\beta=0$; (b) $\alpha=0$, $\beta=3$; and (c) $\alpha=3$ , $\beta=3$.}
\label{fig3}
\end{center}
\end{figure}

\begin{figure}[t!]
\begin{center}
\includegraphics[width=0.98\textwidth]{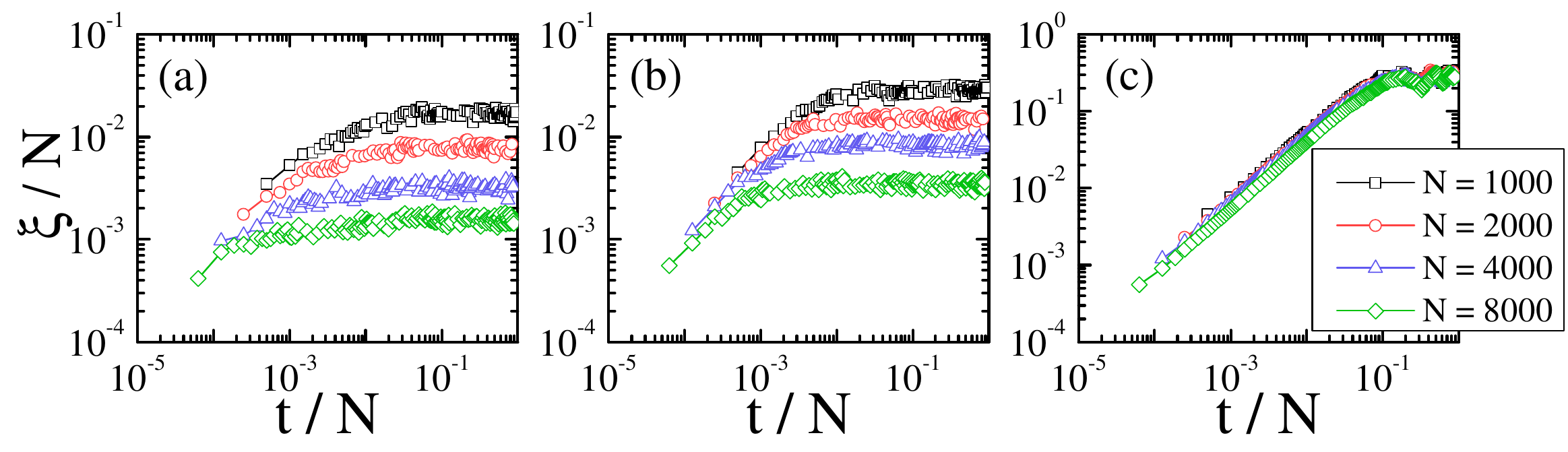}
\caption{ Rescaled participation number ($\xi/N$) versus rescaled time ($t/N$) for (a) $\alpha=3$, $\beta=0$; (b) $\alpha=0$, $\beta=3$; and (c) $\alpha=3$ , $\beta=3$.}
\label{fig4}
\end{center}
\end{figure} 
 
Figures \ref{fig3} and \ref{fig4} 
show a summary of our calculations for the time-dependent spread and participation number 
for an initial delta-like state prepared at the $(N/2)$th site, that is $f_n(0)=\delta_{n,N/2}$ . Those coefficients 
at a later time are evaluated through Eq. (\ref{fnt}) for $N=1000$
up to $8000$ for various combinations of $\alpha$ and $\beta$. 
For comparison purposes,
time and functions of interest were rescaled by the system size $N$.  
We computed $f_n(t)$ until a stationary state could be reached after multiple reflections of the wave packet on the lattice boundaries. 
Therefore, for $\alpha$ and $\beta$ larger than 2 [see Figs. \ref{fig3}(c) and \ref{fig4}(c)] we obtained a 
sharp curve collapse 
thus implying that the wave packet spreads ballistically before reaching the boundaries of the chain.
For $\alpha$ \textit{or} $\beta$ less than $2$, on the other hand, panels (a) and (b) of Figs. \ref{fig3} and \ref{fig4}
there is clearly no collapse, thus
suggesting  a much slower electronic dynamics along the chain~\cite{mauricio2d}.
 
In general lines, our results show that chains with correlated disorder in both diagonal and off-diagonal terms 
can only support the presence of extended states 
once both sources of disorder display strong enough correlations, that is $\alpha, \beta > 2$.
Still, it is very impressive that two competing and independent sources of noise allow for 
coherent transmission of electronic excitations through the chain.
That could, for instance, 
find many applications in quantum communication protocols \cite{bose03,apollarorev}. 
Now, we 
evaluate the robustness of a quantum-state transfer scheme \cite{wojcik05,wojcik07} against our generalized 
disorder model. 

First, let us make further assumptions towards the configuration of the system. We now consider a chain
made up by $N+2$ sites [described by the very same Hamiltonian in Eq. (\ref{hamilcompleto}) 
now with $N \rightarrow N+2$], such that the first and last one will act as, respectively, sender and receiver parties.
For those, particularly, we set $\epsilon_{1} = \epsilon_{N+2} = 0$ and $J_{1} = J_{N+1} = g$
that is, disorder is only present along the communication channel itself (sites $2$ to $N+1$). 
The transfer scheme is based on the weak-coupling model \cite{wojcik05, wojcik07} -- usually worked out in the context of spin chains -- 
where $g$ is set several orders of magnitude weaker than the energy scale of the channel. That forces both end 
sites 
to span their own subspace, with a couple of eigenstates taking the form $\ket{\psi^{\pm}} \approx (\ket{1}\pm\ket{N+2})/\sqrt{2}$ , so that state transmission takes place via coherent dynamics between them. 
Naturally, nearly-perfect transmission shall be expected in ordered chains. 
If that is not the case, the presence of generalized disorder breaks down the mirror and particle-hole symmetries of the system 
thus damaging the effective two-body coupling between the ends of the chain \cite{wojcik07}. 
 
We are now about to show that a high-fidelity quantum-state transfer protocol can actually 
be realized in the presence of correlated fluctuations, involving the whole channel. 
Let us outline the transfer protocol following the original proposal from Ref. \cite{bose03}.
Suppose that Alice wishes to send an arbitrary qubit $\ket{\varphi}_{1} = a\ket{0}_{1}+b\ket{1}_{1}$
to Bob, where $\ket{0}_{i}$ ( $\ket{1}_{i}$) denotes the absence (presence) of an electron at site $i$. 
Then, she arranges for an initial state of the form $\ket{\Psi(0)} = \ket{\varphi}_{1}\ket{0}_{2}\ldots\ket{0}_{N+2}$. 
By letting the system evolve following its natural Hamiltonian dynamics, 
she expects, in the best-case scenario, to have $\ket{\Psi(\tau)} = \ket{0}_{1}\ket{0}_{2}\ldots\ket{0}_{N+1}\ket{\varphi}_{N+2}$
so Bob can properly retrieve the qubit.
A measure for the figure of merit of the protocol
can obtained by averaging the input fidelity over the whole Bloch sphere (for details, see \cite{bose03}):
\begin{equation}\label{avF}
F(t) = \frac{1}{2}+\frac{|c_{N+2}(t)|}{3}+\frac{|c_{N+2}(t)|^2}{6},
\end{equation}
which is basically a monotonic function of the transition amplitude 
between sender and receiver sites, 
$c_{N+2}(t)\equiv \sum_j f_{1}^j f_{N+2}^je^{-iE_jt}. $ [cf. Eq. (\ref{fnt})].
%

\begin{figure}[t!]
\begin{center}
\includegraphics[width=0.9\textwidth]{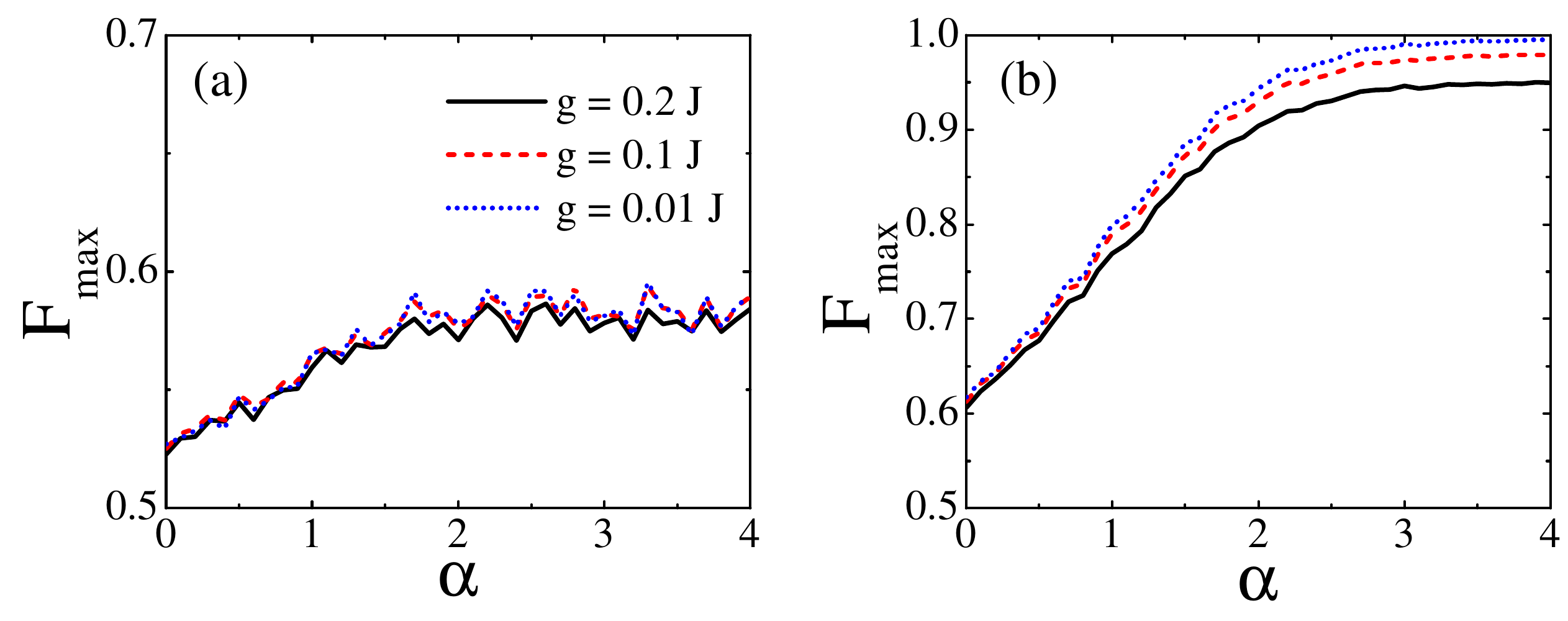}
\caption{Maximum fidelity versus $\alpha$ averaged over 500 independent realizations of disorder
in a 50-site channel (52 sites total) for (a) $\beta=0$ and (b) $\beta=3$. 
Solid, dashed, dotted lines display results for $g/J = 0.2$, $0.1$, $0.01$, respectively.
$F_{\mathrm{max}} \equiv \mathrm{max}\lbrace F(t) \rbrace$ was evaluated over time window $tJ \in [0,2\times 10^{5}]$.}
\label{fig5}
\end{center}
\end{figure}

Here we are concerned with the maximum fidelity $F_{\mathrm{max}} = \mathrm{max}\lbrace F(t)\rbrace$ achieved during a given
interval since the dynamics time scale of the system varies considerably sample by sample. In particular,  we evaluated 
$F_{\mathrm{max}}$ over $tJ \in [0,2\times 10^{5}]$ for about 500 independent realizations of disorder and averaged them out
for every system configuration as shown in Fig. \ref{fig5}.
There, it is clear that an efficient transfer protocol
can be performed through our noisy channel once supported by prominent intrinsic correlations in  
both sources of disorder [see Fig. \ref{fig5}(b)]. We observe that $F_{\mathrm{max}}$ tends 
to saturate after
$\alpha>2$, thus pointing out the crucial role of extended states in the process.
We also highlight in Fig. \ref{fig5}(b)
that we are able to achieve nearly perfect fidelities provided $g$ 
is weak enough, in order to avoid mixing between the channel and sender/receiver subspaces. 

What is most impressive in the results shown above is that even though 
the noisy channel must be augmented with 
strong long-range correlations in order to establish successful quantum-state transfer rounds 
we must point out the fact 
that considerable amounts of disorder are \textit{still} present in the system.
That ultimately destroys the mirror and particle-hole symmetries of the spectrum \cite{yao11} and so, intuitively, 
it should not allow for an effective \textit{resonant} interaction between the outer ends of the chain. Fortunately, it actually does. 
A very useful picture of this can be put forward 
by writing down the sender/receiver decoupled Hamiltonian with renormalized parameters 
obtained through second-order perturbation theory in $g$ [(for details, see Ref. \cite{wojcik07}),
$H_{\mathrm{eff}}= h_{1}\ket{1}\bra{1} + h_{N+2}\ket{N+2}\bra{N+2} - J'\ket{1}\bra{N+2}+\mathrm{h.c.}$,
where 
\begin{align}
h_{1} &= - g^{2}\sum_{k}|f_{2}^{k}|^2/E_{k}, \label{hs} \\
h_{N+2} &= - g^{2}\sum_{k}|f_{N+1}^{k}|^2/E_{k}, \label{hr} \\
J' &= g^{2}\sum_{k}f_{2}^{k}f_{N+1}^{k \, * }/E_{k}, \label{Jeff} 
\end{align}
with the sum in $k$ running over the normal modes of the channel only. 
Recalling that sites $1$ (sender) and $N+2$ (receiver) are tuned to the middle of the band, 
$\epsilon_{1} = \epsilon_{N+2} = 0$, the existence of delocalized states
at this region of the spectrum provided the degree of correlations $\alpha$ and $\beta$ 
are high enough (that is, greater than 2) is such that 
it \textit{masks} the overall asymmetric nature of the chain yielding rather 
balanced distributions of amplitudes $f_{2}^{k}$ and $f_{N+1}^{k}$. Hence, $h_{1}\approx h_{N+2}$
what triggers an effective two-site dynamics with negligible local impurities.  
Moreover, since the renormalized parameters  [Eqs. (\ref{hs}) through (\ref{Jeff})] scales with $E_{k}^{-1}$, 
the outskirts of the band, filled by localized-like states (thus more spatially asymmetric),
have a much weaker influence on them.

\section{Conclusions}

In this work we considered an electronic tight-binding
chain with correlated disorder in both diagonal and off-diagonal terms of the Hamiltonian. 
The fractional Brownian motion was used to generate each corresponding disorder distributions.
We analyzed the localization properties of the system, accounted by the participation ratio of 
its entire spectrum, and also evaluated the electronic dynamics profile along the chain. 
We showed the model supports extended states only if 
both sources of disorder
contain strong intrinsic long-range correlations, for at least $\alpha, \beta > 2$. 
We also investigated a possible application for this class of chains
in the context of quantum-state transfer protocols. By perturbatively coupling
both communicating parties to the noisy chain, it is possible 
to transmit an excitation from one end of the chain to another with very high fidelities as long as
a proper set of delocalized states is available in the spectrum
in order to overcome the spatial asymmetry
induced by disorder.

By tackling the properties of a standard electronic hopping model augmented with twofold long-range-correlated disorder, 
we set the ground for further studies along that direction involving other classes of many-body interacting models. 
Moreover, we also highlight the importance of investigating special types of disorder 
that might occur in solid-state devices for quantum information processing tasks \cite{almeida17-1}.
%

\section{Acknowledgments}

This work was partially supported by CNPq (Grant No. 152722/2016-5),
CAPES, and FINEP (Federal Brazilian Agencies), CNPq-Rede Nanobioestruturas, 
and FAPEAL(Alagoas State Agency).

\end{document}